\documentclass[reviewcopy]{elsart}

\usepackage{graphics}
\usepackage{amssymb}

\newcommand{\latin}[1]{\textit{#1}}
\newcommand{\unit}[1]{\textsl{#1}}
\newcommand{\chem}[1]{\textit{#1}}
\newcommand{\bra}[1]{\ensuremath{\langle#1|}}
\newcommand{\ket}[1]{\ensuremath{#1\rangle}}

\begin{document}
\begin{frontmatter}

\title{QUANTUM CONFINEMENT IN CdSe NANOCRYSTALLITES}
\author{K. E. Andersen},
\author{C. Y. Fong},
\author{W. E. Pickett}
\address{Department of Physics, University of California, Davis,
CA 95616-8677, USA}

\begin{abstract}
Quantum confinement increases the spacing between energy levels as the
nanocrystallite size is decreased. Its qualitative features hold both for
states localized near the center of a nanocrystallite and those near
the surface, such as states due primarily to dangling bonds. However,
different quantitative features are expected because of the different
size constraints on each of these states. Since the majority of atoms
in a typical nanocrystallite are on the surface, contrasting
confinement effects between these two types may prove useful in
predicting how surface state dependent properties, such as optical
absorption, change with the size of the nanocrystallite. By applying
first principles pseudopotential methods to indium doped, uncapped
\chem{CdSe} nanocrystallites containing 17 and 34 atoms, we identify
center and surface localized states. Using the lowest occupied energy
state as a reference, the energy of a state localized near the center
is found to increase 24 \unit{mRy} from the 34 to 17 atom
nanocrystallite. An equivalent surface state within the two cases
studied is not found, but the energy level spacing is speculated to
increase on the order of 100 \unit{mRy} between the 34 and 17 atom
cases based on states that are highly local to the surface, but not
equivalent. Furthermore, we find it's necessary for the impurity to
sit at the center of the nanocrystallite in order for the impurity
states to be electrically active.
\end{abstract}


\end{frontmatter}


\section{Introduction}
The origin of quantum confinement in so-called zero-dimensional
nanocrystallites, such as quantum dots (QDs), is understood to arise
from the spatial confinement of electrons within the crystallite
boundary. It leads to a larger spacing between energy levels as the
size of the nanocrystallite is decreased. Qualitatively this effect is
analogous to the problem of a particle in a box, and efforts to
quantify confinement effects have been the topic of considerable
research\cite{Yof93}. In \chem{CdSe} semiconductor QDs, an important
consequence of quantum confinement is the increase in the band gap as
the QD size is decreased. Since this is observed as an increase in the
energy of the lowest exciton peak as the radius of the QD is
decreased\cite{NERB96,RTYR95}, research in this area has focused
almost exclusively on understanding the energy spectrum of an exciton
as a function of QD radius\cite{Yof93,WZ96,LW99} in order to predict
the optical properties of \chem{CdSe} QDs of an arbitrary
size. However, in principle, quantum confinement should affect every
electronic state within the QD, but \emph{not} equally. For instance,
electrons within a nanocrystallite can be confined in different
spatial regions, such as near the center or surface, and these
different regions should lead to discernible differences in how the
spacing between energy levels changes with respect to the size of the
QD.

Using first principles self-consistent pseudopotential methods, we
quantify these differences for the electronic states within small,
uncapped \chem{CdSe} QDs containing 17 and 34 atoms ($\sim$1 \unit{nm}
diameter). Doped \chem{CdSe} quantum dots are examined, with indium (a
donor) substituted for cadmium, allowing distinct impurity states to
be identified. We contrast confinement effects for impurity states
confined near the center of the QD to those confined near the
surface. Furthermore, we find that placing the impurity at the center
of the QD is necessary for these impurity states to be electrically
active.

Contrasting confinement effects between states localized near the
center to those localized near the surface is important, since the
majority of atoms within a QD are located on the surface. Quantifying
quantum confinement in nanocrystallites, such as QDs, requires
refinement of the particle in a box picture to incorporate the
different ``boxes'' available within a nanocrystallite; in particular
the difference between the surface and center.

\section{Models}
\label{smodels}
\begin{figure}[p]
\includegraphics{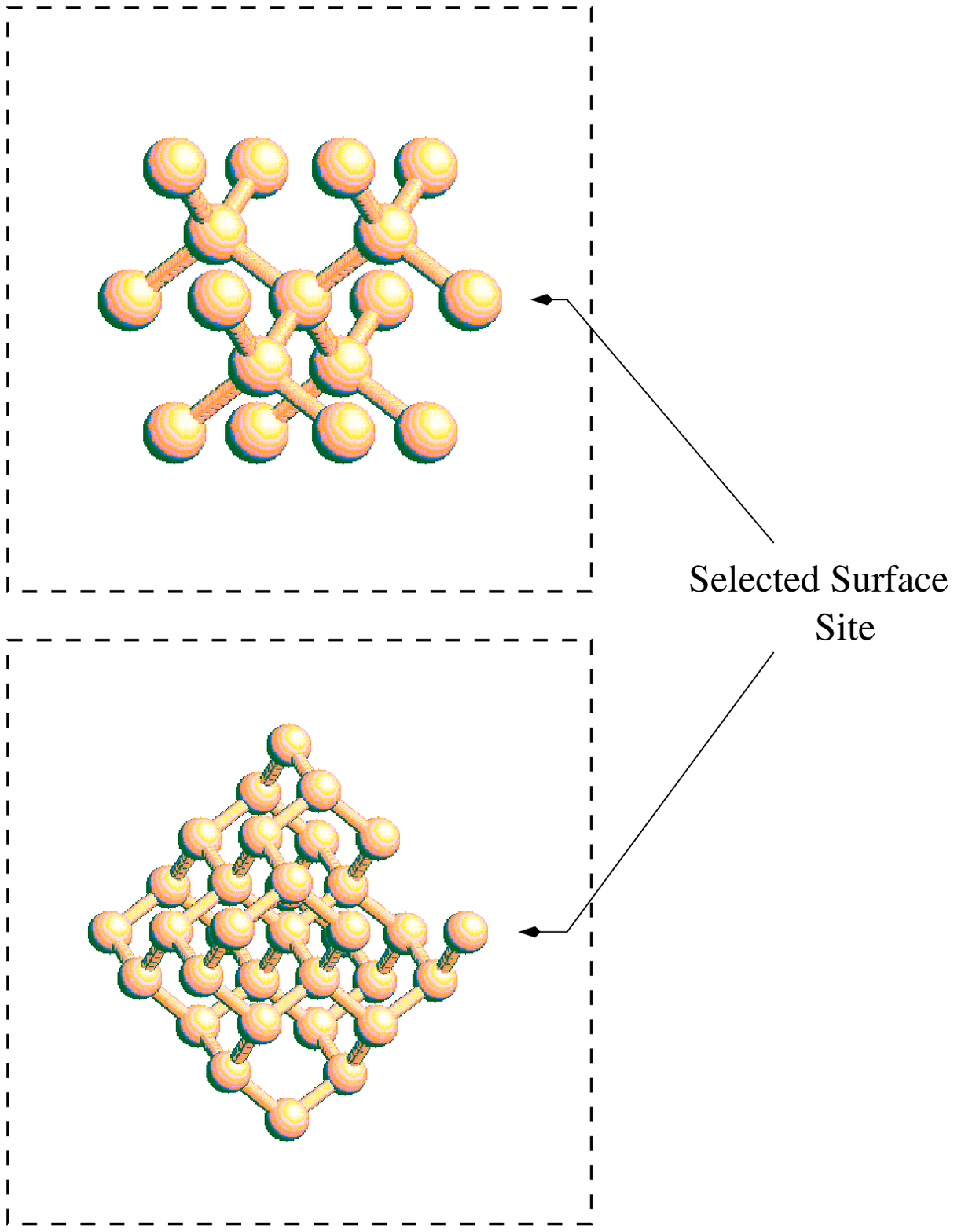}
\caption[Models of \chem{CdSe} quantum dots containing 17 (top) and 34
(bottom) atoms. (Not to scale.) Both quantum dots have a diameter of
approximately 1 \unit{nm} and are centered around a cadmium atom. The
dashed line illustrates schematically the 30.0 and 40.0 \unit{a.u.}
supercell, for the 17 atom and 34 atom cases respectively. Indium is
substituted for cadmium at either the center or the surface site
shown.]{} 
\label{models}
\end{figure}

Two sizes of \chem{CdSe} QDs containing 17 and 34 atoms were modeled
(Fig. \ref{models}). These were constructed using the cubic zincblende
crystal structure up to a cutoff radius of 8.35 and 11.81 \unit{a.u.},
for the 17 and 34 atom cases respectively, from a central \chem{Cd}
atom. This structure was then surrounded within a supercell of 30.0
and 40.0 \unit{a.u.} respectively, on each side. No relaxation of the
ionic positions was done, since such relaxation can itself be a
formidable task and the effect we are studying should not depend
strongly on it.

These models represent uncapped, non-interacting collodial \chem{CdSe}
quantum dots with diameters of roughly 1 \unit{nm}. To make the
confining region at the surface similar between the two cases, an atom
was removed from the 34 atom QD to make the bonding environment at the
selected surface site more like that of the 17 atom case (shown in
Fig. \ref{models}). Without this atom, both the 17 and 34 atom cases
have a site on the surface that is missing three bonds. Indium was
then substituted for cadmium at either this site or the center.

\section{Methods}

\subsection{Electronic Properties}
Norm-conserving pseudopotentials of the Hamann form\cite{Ham89}
were used in conjunction with the local density approximation of
density functional theory\cite{HK64,KS65b} to calculate all
electronic properties. The wave function was expanded in a plane
wave basis with a 50.0 \unit{Ry} energy cutoff, using
approximately 150,000 and 375,000 plane waves for the 17 and 34
atom cases respectively.

\subsection{Identifying States}
The added electron associated with the indium atom introduces a
unique impurity state that, when substituted at either the center
or a surface site, provides a way to identity a particular state
between QDs of different size that will be localized either near
the center or surface. By calculating self-consistently the wave
function and energy spectrum using the local density approximation
of density functional theory it is possible to 1) identify these
impurity states by analyzing the projected density of states local
to an atom and 2) quantify the confinement effect on these states,
using the lowest energy state as a reference.

The projected density of the $n$th state $D_{n}(\epsilon, r,
\vec{\tau})$ local to an atom at the location $\vec{\tau}$ is used to
identify states.
\begin{eqnarray}
\label{dos}
D_{n}(\epsilon, r, \vec{\tau}) & = & \frac{1}{V} \sum_{l=0}^{\infty} \sum_{m = -l}^{l} | \bra{lm} \ket{\psi_{n}(\vec{r} - \vec{\tau})}_{ang} |^{2}\ \delta(\epsilon - \epsilon_{n}) \\
                               & = & \frac{1}{V} \sum_{l=0}^{\infty} \sum_{m = -l}^{l} |4\pi \sum_{\vec{G}} a_{\vec{G}, n} \e^{-i \vec{G}\cdot\vec{\tau}} j_{l}(G r) Y_{lm}^{\ast}(\theta_{\vec{G}}, \phi_{\vec{G}}) |^{2} \ \delta(\epsilon - \epsilon_{n}) \nonumber
\end{eqnarray}
Here $\psi_{n}$ and $\epsilon_{n}$ are the wave function and energy of
the $n$th state, and $V$ is the volume of the unit cell.
$\psi_{n}(\vec{r} - \vec{\tau}) = \sum_{\vec{G}} a_{\vec{G}, n}
\e^{i\vec{G} \cdot (\vec{r} - \vec{\tau})}$ is projected onto the
spherical harmonic basis $|\ket{lm}$. Based on the magnitude of the \chem{s}
($l=0$), \chem{p} ($l=1$), and \chem{d} ($l=2$) components it is
possible to identify the states associated with an atom, in
particular, the impurity.

\section{Results}
\begin{figure}[p]
\includegraphics{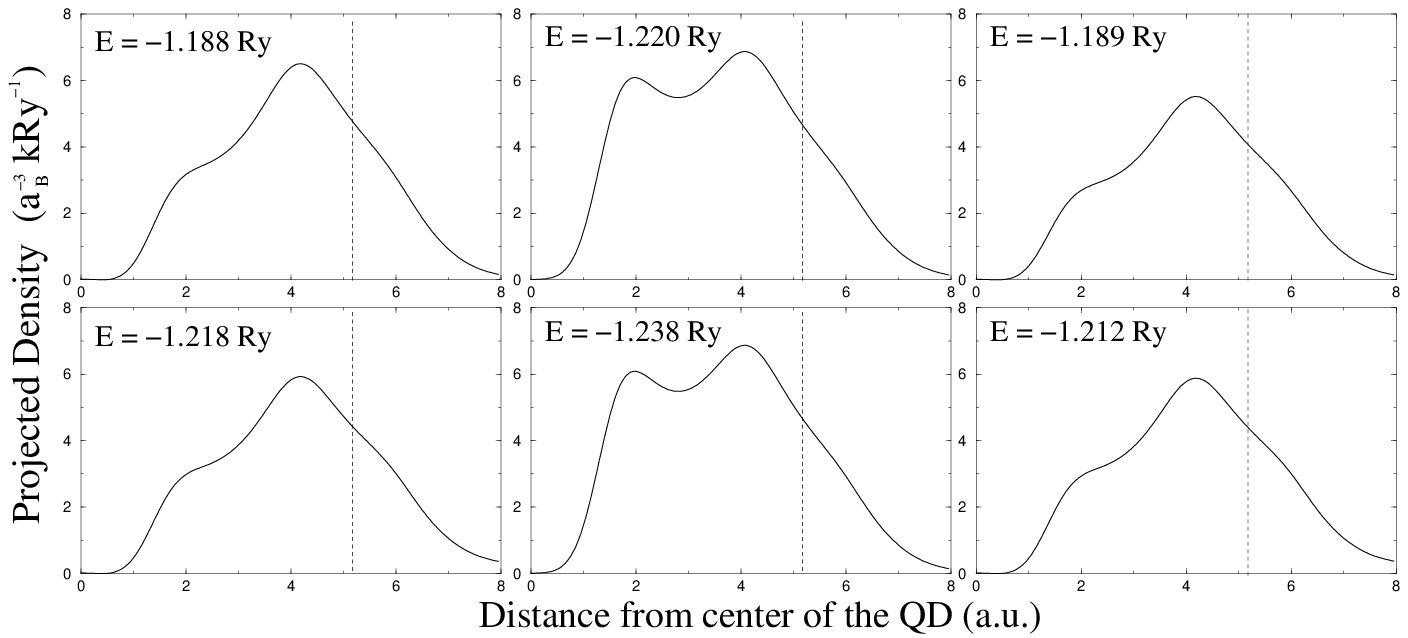}
\caption[Projected density for the lowest energy state, with the
projection origin at the center of the QD. The columns, from left to
right, are for the case with no impurity, the impurity at the center,
and the impurity at the surface for the 17 atom QD (graphs along the
top row) and 34 atom QD (bottom row). The location of the neighboring
\chem{Se} atom is shown by a line at 5.11 \chem{a.u.}. The energy of
the state is labeled.]{}
\label{reference}
\end{figure}

The lowest energy state is used as a reference to compare energies
between QDs of different size and impurity location. This state, which
can be thought of as originating from the \chem{4s} state of the
interior shell of \chem{Se} atoms, was selected because it is expected
to be the most inert and therefore least affected by the addition of
the impurity atom. Fig. \ref{reference} shows the projected density,
as defined in Eq. \ref{dos}, of the lowest energy state for the cases
considered. The projection origin $\vec{\tau}$ is the center of the
QD, which corresponds to either a \chem{Cd} or \chem{In} atom. With
the projection oriented as such, the \chem{s} character shown is a
result of a symmetric combination of orbitals on each of the four
neighboring, tetrahedrally positioned \chem{Se} atoms. No substantial
\chem{p} or \chem{d} components are present for this state. For the
case of \chem{In} at the center of the QD (the middle column of
graphs), the more attractive \chem{In} pseudopotential (compared to
\chem{Cd}) distorts these orbitals leading to an increase in the peak
between the \chem{In} and \chem{Se} atoms at approximately 2
\unit{a.u.}.

The assumption of the lowest energy state being inert holds for the
case of the impurity at the surface of the quantum dot, which only
differs in energy from the lowest energy state with no impurity by 0.7
and 5.7 \unit{mRy} for the 17 and 34 atom cases. To within the
accuracy of our calculations, which are on the order of 1 \unit{mRy},
these two states are essentially identical in energy.

However, the assumption doesn't hold when the impurity is at the
center.  The energy difference between the lowest energy state with
the impurity at the center and the lowest energy state with no
impurity is 31.7 \unit{mRy} for the 17 atom case and 20.2 \unit{mRy}
for the 34 atom case. We attribute this non-constant shift in energy
of the lowest energy state between QDs of different size to quantum
confinement on the lowest state. For the impurity at the surface, the
effect of quantum confinement is negligible; however, when the
impurity is at the center of the QD there is a 11.5 \unit{mRy}
discrepancy in the shift of the lowest energy state.

With the impurity at either the center or a surface site, the
projected density is analyzed to identify an analogous state between
the 17 and 34 atom quantum dots. The 34 atom case, because it has more
electrons, necessarily has more electronic states. However, if a state
is found within its energy spectrum that is similar in character to a
state in the energy spectrum of the 17 atom QD, then these states can
be considered to be physically the same. Comparing the energies of
such a state with respect to a suitable reference will then quantify
the effect due to confinement.

\section{Discussion}

\begin{figure}
\includegraphics{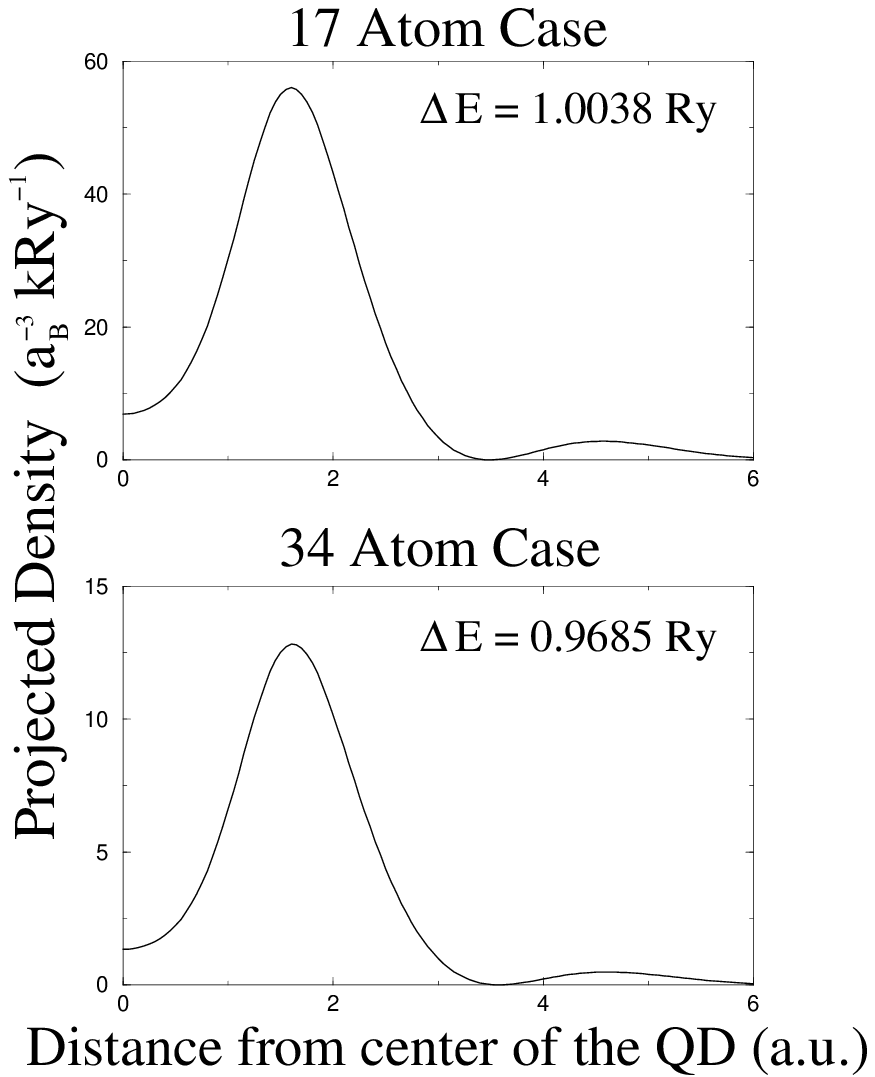}
\caption[The projected density for two equivalent states in the 17
(top) and 34 (bottom) CdSe quantum dot models. These states, at the
lowest unoccupied (17 atom case) and highest occupied (34 atom case)
energy states, are localized spatially near the center of the quantum
dot. The relative energy of the state is given with respect to the
lowest occupied energy level of the system.]{}
\label{center}
\end{figure}

For the case of the impurity at the center (Fig. \ref{center}), it was
possible to identify a unique state between the two sizes of QDs. For
the 17 atom case, this state was the lowest unoccupied energy state
and is 1.0038 \unit{Ry} above the lowest energy state of the 17 atom
QD. For the 34 atom case, it was the highest occupied energy state at
an energy of 0.9685 \unit{Ry} above the lowest state. Na\"{\i}vely,
this suggests that quantum confinement has increased the energy level
spacing for this state by 35 \unit{mRy} as the size of the QD was
reduced from 34 to 17 atoms. However, because the lowest energy state
of the 17 atom case has itself been affected by confinement this value
should instead be 24 \unit{mRy}.

\begin{figure}[p]
\includegraphics{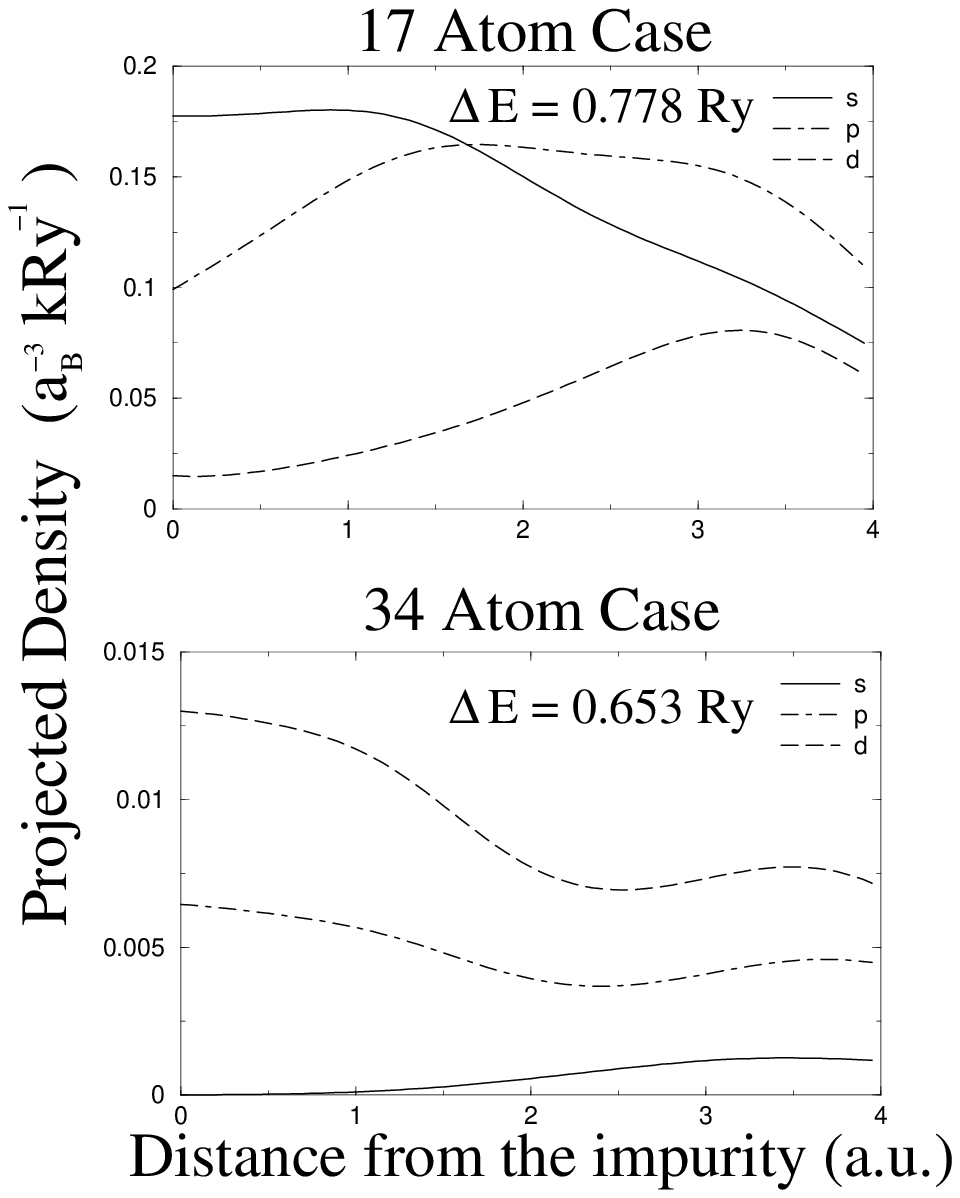}
\caption[Projected density for states that are localized near the
impurity atom (\chem{In}) at the surface of the 17 (top) and 34
(bottom) atom QDs. Both of these states are lower in energy than the
highest occupied energy level. The relative energy of the state is
given with respect to the lowest occupied energy level of the system.]{}
\label{surface}
\end{figure}

With the impurity at the surface, finding a similar state between the
two sizes of QDs is complicated by the surface; even \emph{with} the
removal of one atom from the surface to make the impurity sites
similar, as discussed in \S\ref{smodels} and shown in
Fig. \ref{models}. In contrast to the case of the impurity at the
center, no states near the highest occupied energy state were
significantly localized near the impurity atom. However, there were
significantly localized states below this energy level. An example of
two such states is shown in Fig. \ref{surface}. Physically, the
impurity state at the surface is more bound than the corresponding
impurity state at the center. At the center the bonding with the
nearest neighbor layer of \chem{Se} is complete with
\unit{Cd}. Substituting \chem{In} for \chem{Cd} introduces an extra
electron that is relatively unlocalized. This is supported by the
predominantly \chem{s} projected density seen in Fig. \ref{center}. At
the surface, however, the nearest neighbor layer of \chem{Se} has only
three neighboring \chem{Cd} atoms, instead of four. The extra electron
introduced when \chem{In} is substituted for \chem{Cd} therefore
participates in bonding, resulting in a lower energy for that state.

Since the identification of a unique impurity related surface state
for both sizes of QD isn't possible, we proceed by comparing two
states which are localized near the impurity atom, but not
equivalent. The projected density of the chosen states is in
Fig. \ref{surface}. Other states were also found to be localized
around the impurity. The selected states were chosen because they were
the most localized states near the highest occupied energy state. For
the 17 atom case the chosen state is at an energy of 0.778 \unit{Ry}
above the lowest occupied energy state. Within the energy spectrum of
the 34 atom case, the chosen state is at an energy of 0.6530 \unit{Ry}
above the lowest occupied energy state of that system. To compare, the
spacing between the energy level of the chosen state and the lowest
occupied energy state has increased on the order of 100 \unit{mRy}
between the 34 and 17 atom QDs.

The observation that confinement effects at the surface are more
pronounced than in the center of the QD is suggestive, but hardly
convincing based on the \latin{ad hoc} assumption made that the
localized states chosen could be compared between the two sizes of
QDs. In ongoing research we are investigating the charge density of
these states to see if a more certain identification can be made. In
addition, we're looking at a 71 atom QD model (essentially another
layer of \chem{Se} and \chem{Cd}) that mimics the surface bonding
environment of the 17 atom case much more closely, with the hope of
eliminating this ambiguity.

\section{Conclusion}

In identifying a reference energy, which is a necessary prerequisite
of any quantitative analysis of quantum confinement, we encountered
the complication that the most suitable energy level, the lowest
occupied energy state, was itself affected by quantum confinement. By
making a comparison to the case with no impurity, it was possible to
quantify this discrepancy to find that the energy level spacing for
states localized near the center of the QD increased 24 \unit{mRy}
between the 34 and 17 atom cases. For states localized near the
surface, although identifying a suitable reference state was
straightforward, unambiguously identifying a state near the surface in
both the 17 and 34 atom cases was impossible due to complications at
the surface. Our results suggest that the energy level spacing of such
a state would increase on the order of 100 \unit{mRy} in going from
the 34 to the 17 atom QD. Furthermore, in identifying impurity states
we found that doping at the center was necessary to introduce
electrically active impurity states. Such states, either at or just
above the highest occupied energy level, would be expected to
contribute to transport and optical properties within these systems.

\begin{ack}
KEA acknowledges support from the UC Davis IGERT Program on
Nanophases, NSF Grant No. DGE-9972741. CYF acknowledges support from
NSF Grant No. INT-9872053.
\end{ack}

\bibliography{france}
\bibliographystyle{elsart-num}

\newpage

\listoffigures

\end{document}